\documentclass[11pt,letter]{emulateapj}
\slugcomment{{\sc Submitted to ApJL:} September 22, 2008}
\usepackage{amsmath}
\usepackage{lscape}

\shortauthors{Fuentes, George \& Holman 2008}
\shorttitle{Subaru pencil-beam search for $m_R\sim27$ TNOs}

\def\Eq#1{\rm Eq.~\ref{#1}}
\def\Fig#1{\rm Fig.~\ref{#1}}

\def\km{~\rm km}

\def\au{~\rm AU}

\def\sqdeg{~\rm deg^2}
\def\aph{~\rm ''~h^{-1}}

\begin{document}

\bibliographystyle{apj}

\title{A Subaru pencil-beam Search for $m_R\sim27$ Trans-neptunian bodies\altaffilmark{1}}
\author{Cesar I.\,Fuentes\altaffilmark{2}, Matthew R.\,George\altaffilmark{2,3}, Matthew J.\,Holman\altaffilmark{2} }
\altaffiltext{1}{Based on data collected at Subaru Telescope, which is operated by the National Astronomical Observatory of Japan.}
\altaffiltext{2}{Harvard-Smithsonian Center for Astrophysics, 60 Garden Street, Cambridge, MA 02138, USA; cfuentes@cfa.harvard.edu}
\altaffiltext{3}{Current Address: Department of Astronomy, University of California at Berkeley, Berkeley, CA 94720, USA}

\begin{abstract}
  We present the results of an archival search for Trans-neptunian
  objects (TNOs) in an ecliptic field observed with Subaru in
  2002. The depth of the search allowed us to find 20 new TNOs with
  magnitudes between $R=24$ and $27$. We fit a double power law model
  to the data; the most likely values for the bright and faint power
  law exponents are $\alpha_1$=$0.73_{-0.09}^{+0.08}$ and
  $\alpha_2$=$0.20_{-0.14}^{+0.12}$; the differential number density
  at $R=23$ is $\sigma_{23}$=$1.46_{-0.12}^{+0.14}$ and the break
  magnitude is $R_{eq}$=$25.0_{-0.6}^{+0.8}$.  This is the most
  precise measurement of the break in the TNO luminosity function to
  date. The break in the size distribution corresponds to a diameter
  of $D = 90\pm30$ km assuming a 4\% albedo.
\end{abstract}

\section{Introduction}\label{sec:intro}
The TNO size distribution is the result of the formation and
collisional history of its members. Models of the physical evolution
assume the processes at play result in a characteristic power law size
distribution. In the model of \citet{Pan.2005} of strengthless TNOs,
the collisional evolution of the objects leads to an evolving size
distribution that changes slope at the size at which the collisional
lifetime of an object equals the age of the system. This analytical
result confirms those obtained from numerical
simulations~\citep{Kenyon.1999,Davis.1999,Kenyon.2004}. The precise
location of the break in the size distribution is an important
measurement to link TNO formation and evolution models to the current
TNO population (see \citealt{Kenyon.2008} for a review).

The TNO luminosity function is defined by the distance, albedo, and
size distributions of TNOs, and it is readily observed.  The distance
distribution exhibits a sharp edge at $50\au$ \citep{Trujillo.2001}
that has been observed in other surveys \citep{Gladman.2001,
Fuentes.2008}. Though it is customary to use a 4\% albedo,
correlations with size have been reported (see
\citealt{Stansberry.2008} for a review). After making sensible
assumptions for the albedo and distances of the observed TNOs, the
size distribution can be derived from the luminosity function. Several
searches have been completed (see \citealt{Bernstein.2004,
Fuentes.2008} for a review), the deepest being the search by
\citet{Bernstein.2004} with the {\it Hubble Space Telescope}
(HST). While previous results sampled the size distribution up to
magnitude $R=26$ and were consistent with a power law luminosity
function, the \citet{Bernstein.2004} survey found a significant
underabundance of objects at $R\sim28$, compared to that predicted
from extrapolating the luminosity function at brighter
magnitudes. They concluded a break in the luminosity function must
occur between $R\sim25$ and $28$.

Since \citet{Bernstein.2004} announced their results, a number of
ground-based surveys for fainter TNOs have been conducted in an effort
to bridge the divide between the brighter ($R$$\sim$$23$) TNOs found
through wide-field searches such as the Deep Ecliptic
Survey~\citep{Elliot.2005} and the Canada-France Ecliptic Plane
Survey~\citep{Jones.2006} and those found by \citet{Bernstein.2004}.
\citet{Fuentes.2008} discovered 82 TNOs in an archival search of
Subaru data.  This survey reached a limiting magnitude of $R=25.7$ and
successfully detected the break in the luminosity function. In the
present work we build upon that earlier work, increasing the limiting
magnitude of ground-based surveys to $R\sim27$ and further narrowing
the gap between ground-based and space-based TNO surveys. Constraints
on even fainter objects have been placed by stellar occultation
surveys \citep{Bickerton.2008, Zhang.2008}.

We present the details of the data in the next
section. \S~\ref{sec:mod} outlines the method used to analyze and
search for TNOs in the data. The characterization of the search using
a synthetic population is presented in \S~\ref{sec:effic}. In
\S~\ref{sec:results} and \S~\ref{sec:conc} we present our results and
discuss what they imply for the formation and evolution of the TNO
population.

\section{Data}\label{sec:data}
The apparent motion of solar system bodies limits the useful exposure
time to the period required for an object to move a PSF width; longer
exposures spread the signal along a trail and, correspondingly,
increase the contribution of the sky background.  To overcome this, we
consider a series of short exposures and shift the successive images
to compensate for the motion of the object.  In this way the signal
from the source can dominate the noise from the background.  This
``pencil beam'' or ``digital tracking'' approach has been successfully
applied in searches for TNOs and outer planet satellites
\citep{Allen.2001, Gladman.2001, Holman.2004, Kavelaars.2004,
Fraser.2008}.

Using Subaru's electronic archive SMOKA~\citep{Ichikawa.2002}, we
identified a data set well suited to such a search, a series of 148
consecutive $120~\mathrm{s}$ exposures of a single, ecliptic field
observed with Subaru on UT 2002 September 2 over the course of
8~hours.  These data were collected with
Suprime-Cam~\citep{Miyazaki.2002}, a 10-CCD mosaic camera with a $34'
\times 27'$ field of view and an image scale of
$0\farcs202~\mathrm{pix}^{-1}$.  It is our understanding that these
data were originally collected for the Subaru Main Belt Asteroid
Survey (see~\citealt{Yoshida.2007}).  The field, R.A. 22:41:38, Dec
-07:37:35, is less than $1^{\circ}$ from the ecliptic and was observed
near opposition using the ``Cousins R'' filter. Although brief
intervals of clouds are apparent from the images, the night was mostly
photometric, with stable seeing of $\sim0\farcs7$.

We trimmed, bias subtracted and flat divided the images with
calibrations obtained on the same night and the next one using
standard IRAF\footnote{IRAF is distributed by the National Optical
Astronomy Observatories, which are operated by the Association of
Universities for Research in Astronomy, Inc., under cooperative
agreement with the National Science Foundation.} routines.

\section{Moving Object Detection}\label{sec:mod}
We determined an astrometric solution for each individual CCD using
the 2MASS point source catalog~\citep{Cutri.2003}. The RMS of the
solution was $\sim0\farcs2$, comparable to that of the catalog and
much smaller than the typical seeing. We then registered every image
to the same astrometric reference.

Next, we inserted a population of synthetic objects that will be later
used to determine the efficiency of our search (the details of this
process are given in \S~\ref{sec:effic}).

To detect TNOs moving at any physically plausible velocity,
we defined a grid of rates, parallel and perpendicular to the
ecliptic. A TNO at a typical distance of $42\au$ exhibits a
parallactic motion of $\sim3\aph$, moving about 120 pixels over the
length of the observations. We searched rates from $0\farcs7$ to
$5\farcs1~{\rm h^{-1}}$ for the parallel rate to cover the range
between $\sim20$ and 200$\au$, and perpendicular rates in the range
$-1\farcs4$ to $1\farcs 4~{\rm h^{-1}}$. Furthermore, we restricted
our attention to directions of apparent motion within $15 ^ \circ$ of
the ecliptic, which nevertheless permits the detection of highly
inclined TNOs. We searched these rates with a resolution of $0\farcs
1~\rm{h^{-1}}$ along both axes (this resolution ensures that the
signal in the first image can be aligned with that in the last for any
object).

Prior to combining the images, we used the ISIS
package~\citep{Alard.2000} to PSF match and subtract a template from
each individual exposure, thus eliminating any source that is
stationary and of constant brightness.  For each individual CCD, the
template image was the median combination of 10 of the best-seeing
images. Saturated stars were masked at this point to avoid spurious
detections due to imperfect subtraction.

To optimally combine the 148 images we defined a ``weight'' for each
image based on both its sky background and seeing: $w_i= \sigma_i^{-2}
\theta_i^{-2} (\sum_j \sigma_j^{-2} \theta_j^{-2} )^{-1}$, where
$\theta$ is the FWHM and $\sigma$ is the standard deviation of the
background in counts. We defined a reference time to be the weighted
average of the exposure mid-times.  For each of the 736 combinations
of parallel and perpendicular rates we shifted all subtracted images
to the reference time and computed the weighted average of them.

We searched each of the ``shifted and added'' images for point sources
that corresponded to real or implanted objects moving at the
corresponding rate.  For this we used a combination of the SExtractor
package~\citep{Bertin.1996} and a wavelet source detection routine
\citep{Petit.2004}, each with a $3\sigma$ detection threshold.  These
two approaches rely on very different image properties and, thus, have
different noise characteristics.  We have found that the intersection
of the results from these two routines strongly discriminates against
false detections.  Nevertheless, we find $\sim300$ detections per
coadded CCD at each combination of rates parallel and perpendicular to
the ecliptic.

We comb through these with an automated search algorithm.  The
algorithm considers all detections in all of the parallel and
perpendicular rate combinations (roughly $2.2\times 10^5$ detections
per CCD), projects them on the sky plane, and then searches for
clusters of detections.  Assuming each cluster corresponds to a moving
object detected in one or more of the coadded images for rates close
to the real, or synthetic, one, the code selects the single detection
from every cluster that has the smallest ellipticity and largest flux.
We visually inspect the resulting $\sim300$ candidates per CCD in
order to reject any obvious spurious detections; those rejected are
typically poorly subtracted stars that were not adequately masked, or
cosmic rays that resulted in multiple detections. This inspection
process is extremely fast and decreases the number of possible moving
objects to $\sim60$ per CCD.  We further examine each of these
remaining candidates to determine the parallel and perpendicular rates
that yield the PSF that best resembles a normal one. In this way we
visually found the best rate of motion. In this process another 40\%
of the remaining objects were rejected.

The novelty in this method is the initial automated search and visual
filter for candidates that are later checked by blinking through
different rates. The alternatives are a fully automated search
algorithm run at a higher detection threshold to eliminate false
positives~\citep{Bernstein.2004} or a visual search through the whole
field and all possible rates~\citep{Gladman.2001}. We expect that a
fully visual search would be more sensitive, leaving aside the issue
of human fatigue.  To test this we completed a visual search of 20\%
of the area (two of the ten CCDs).  In this test we visually confirmed
or rejected potential candidates identified by the two source
detection algorithms, as was done by \citet{Holman.2004}.  The
comparison showed our approach to be $\sim 10\%$ less sensitive
compared to a visual search, roughly independent of magnitude and
rate.

\section{Control Population and Detection Efficiency}\label{sec:effic}
To calibrate our search we used a control population of objects that
resemble real TNOs both astrometrically and photometrically. We
considered a range of distances ($30-200\au$) and magnitudes
($R=24.5-29.5$), as well as the full range of eccentricities and
inclinations to create a population of synthetic TNOs. Ephemerides for
this population were created using a modified version of the {\it
Orbfit} routines \citep{Bernstein.2000}.

For each CCD $\sim10$ bright stars were used to determine a PSF
model. These stars are also used to account for changes in the seeing
and the atmospheric transparency when implanting objects.

\begin{figure}[ht]
  \epsscale{1.0} \plotone{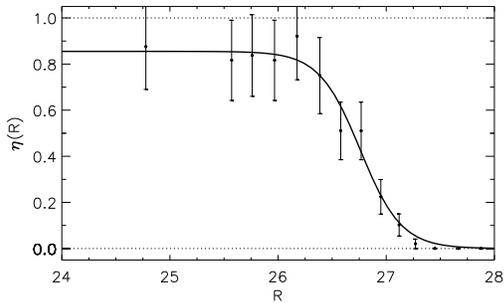}
  \caption{\label{fig:effofmag} Detection efficiency as a function of
    magnitude. The solid line corresponds to $\Eq{eqn:eff}$, with best
    fit values: $A$=$0.86\pm0.07$, $R_{50}$=$26.76\pm0.06$ and
    $w$=$0.38\pm0.04$. }
\end{figure}

We recovered 312 synthetic TNOs and used them to determine the
efficiency as a function of $R$ magnitude. The result is shown in
Figure~\ref{fig:effofmag}.  The detection efficiency is well
represented by the following function:
\begin{equation}\label{eqn:eff}
\eta(R) = \frac{A}{2} \left(1-\tanh{\frac{R-R_{50}}{w}} \right)
\end{equation}
where the best fit values are $A$=$0.86\pm0.07$,
$R_{50}$=$26.76\pm0.06$ and $w$=$0.38\pm0.04$. The maximum efficiency
of our search is 86\%, and it reaches half this value at magnitude
$R=26.76$.

The USNO-B Catalog was used to tie the photometry of our objects to a
standard system. A number of USNO-B stars are unsaturated in our field
($m_R>18$). The photometric uncertainty of the catalog is 0.3 mag
\citep{Monet.2003}. Seven isolated stars were selected, and compared
to the photometry of an exposure from a photometric portion of the
night. The flux $f_5$ for a 5-pixel ($1\farcs01$) aperture over a time
$t$ in seconds was found to be:
\begin{equation}
R =    27.45 - 2.5 \log{f_5 / t} .
\end{equation}
An aperture correction of 0.076 mag is included in this
formula.

\section{Results and Analysis}\label{sec:results}

Our search resulted in 20 new objects, summarized in Table
\ref{tab:objects}. The distances to these objects are estimated using
their parallactic motion, assuming circular orbits. To estimate the
uncertainty in these distance estimates, we compare the precision of
the parallactic rate to that in \citet{Fuentes.2008} and adopt the
uncertainty in those distances ($2.5\au$). There are three objects
with distances closer than 30 AU; given the poor orbital information
we assume these objects are Plutinos and retain them in our sample.

\begin{deluxetable}{lcccccc}
\tabletypesize{\scriptsize}
\tablecaption{\sc  TNO properties\tablenotemark{a}}
\tablewidth{0pt}
\tablehead{
  \colhead{} & 
  \colhead{$R_{mag}$} &
  \colhead{${\rm R.A.}$} &
  \colhead{${\rm Decl.}$} &
  \colhead{$d{\rm R.A.}/dt$} &
  \colhead{$d{\rm Decl.}/dt$} &
  \colhead{${\rm d_{par}}$}
  \\
  \colhead{} &
  \colhead{} &
  \colhead{} &
  \colhead{} &
  \colhead{${\rm (''~h^{-1})}$} &
  \colhead{${\rm (''~h^{-1})}$} &
  \colhead{${\rm(AU)}$}
}
\startdata
sd1  & $26.7$ & 22:40:46.967 & -07:28:56.80 & -3.19 & -1.53 & 35.5 \\
sd2  & $24.0$ & 22:40:38.889 & -07:34:33.23 & -2.99 & -1.23 & 39.2 \\
sd3  & $24.6$ & 22:40:48.201 & -07:33:30.04 & -2.99 & -1.23 & 39.2 \\
sd4  & $26.7$ & 22:42:46.043 & -07:25:05.09 & -4.39 & -1.27 & 26.8 \\
sd5  & $26.6$ & 22:42:29.147 & -07:34:07.96 & -2.80 & -1.15 & 42.1 \\
sd6  & $25.9$ & 22:42:39.499 & -07:34:09.84 & -3.55 & -1.46 & 32.4 \\
sd7  & $25.4$ & 22:42:26.506 & -07:34:58.77 & -2.67 & -1.21 & 43.6 \\
sd8  & $25.1$ & 22:42:31.162 & -07:36:21.54 & -2.71 & -1.12 & 43.6 \\
sd9  & $25.2$ & 22:42:36.655 & -07:28:27.49 & -2.28 & -0.83 & 53.6 \\
sd10 & $25.9$ & 22:42:17.085 & -07:24:26.88 & -3.19 & -1.53 & 35.5 \\
sd11 & $24.6$ & 22:42:18.460 & -07:35:09.91 & -2.89 & -1.19 & 40.6 \\
sd12 & $26.1$ & 22:41:33.852 & -07:39:08.51 & -2.99 & -1.23 & 39.2 \\
sd13 & $26.6$ & 22:41:33.660 & -07:40:31.79 & -4.03 & -1.88 & 27.5 \\
sd14 & $26.8$ & 22:40:57.890 & -07:44:53.89 & -2.89 & -1.19 & 40.6 \\
sd15 & $26.6$ & 22:41:22.837 & -07:42:07.71 & -4.22 & -1.41 & 27.5 \\
sd16 & $25.7$ & 22:41:01.823 & -07:35:49.27 & -2.86 & -1.29 & 40.6 \\
sd17 & $26.6$ & 22:41:09.107 & -07:25:13.78 & -2.78 & -0.93 & 43.6 \\
sd18 & $24.9$ & 22:41:09.797 & -07:26:31.27 & -2.71 & -1.12 & 43.6 \\
sd19 & $26.7$ & 22:41:44.583 & -07:31:58.21 & -4.71 & -1.29 & 24.9 \\
sd20 & $26.8$ & 22:42:38.548 & -07:47:45.67 & -2.44 & -1.23 & 47.2
\enddata
\tablenotetext{a}{ Positions and rates valid for MJD 52519.451992. The
  uncertainties can be found in the text. }
\label{tab:objects}
\end{deluxetable}

\begin{figure}[ht]
  \epsscale{1.0} \plotone{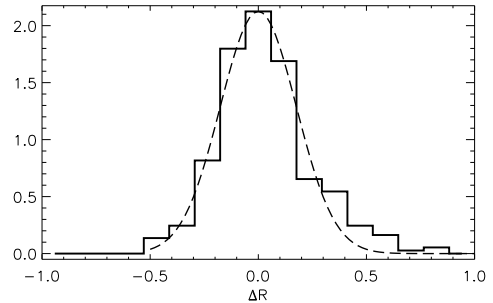}
  \caption{\label{fig:magerr} Histogram of the magnitude difference
    between implanted and measured magnitudes $\Delta R$. The dashed
    line is a Gaussian of width 0.18 mag.  }
\end{figure}

We were able to measure the error in our photometry via comparison of
the implanted and recovered magnitudes, plotted in Figure
\ref{fig:magerr}. This allowed us to set the aperture correction and
include any uncertainty introduced by our method. The net uncertainty
is $\sim$0.2 mag, similar to the expected variance given the sky
background; we adopted it as the 1-$\sigma$ uncertainty for all our
detections. As for the accuracy in the rate of motion parallel and
perpendicular to the ecliptic is $\sim0\farcs1~\mathrm{h}^{-1}$.  The
corresponding uncertainty in the angle to the ecliptic is
$\sim3~\mathrm{deg}$.

\begin{figure*}[ht]
  \epsscale{1.01} \plottwo{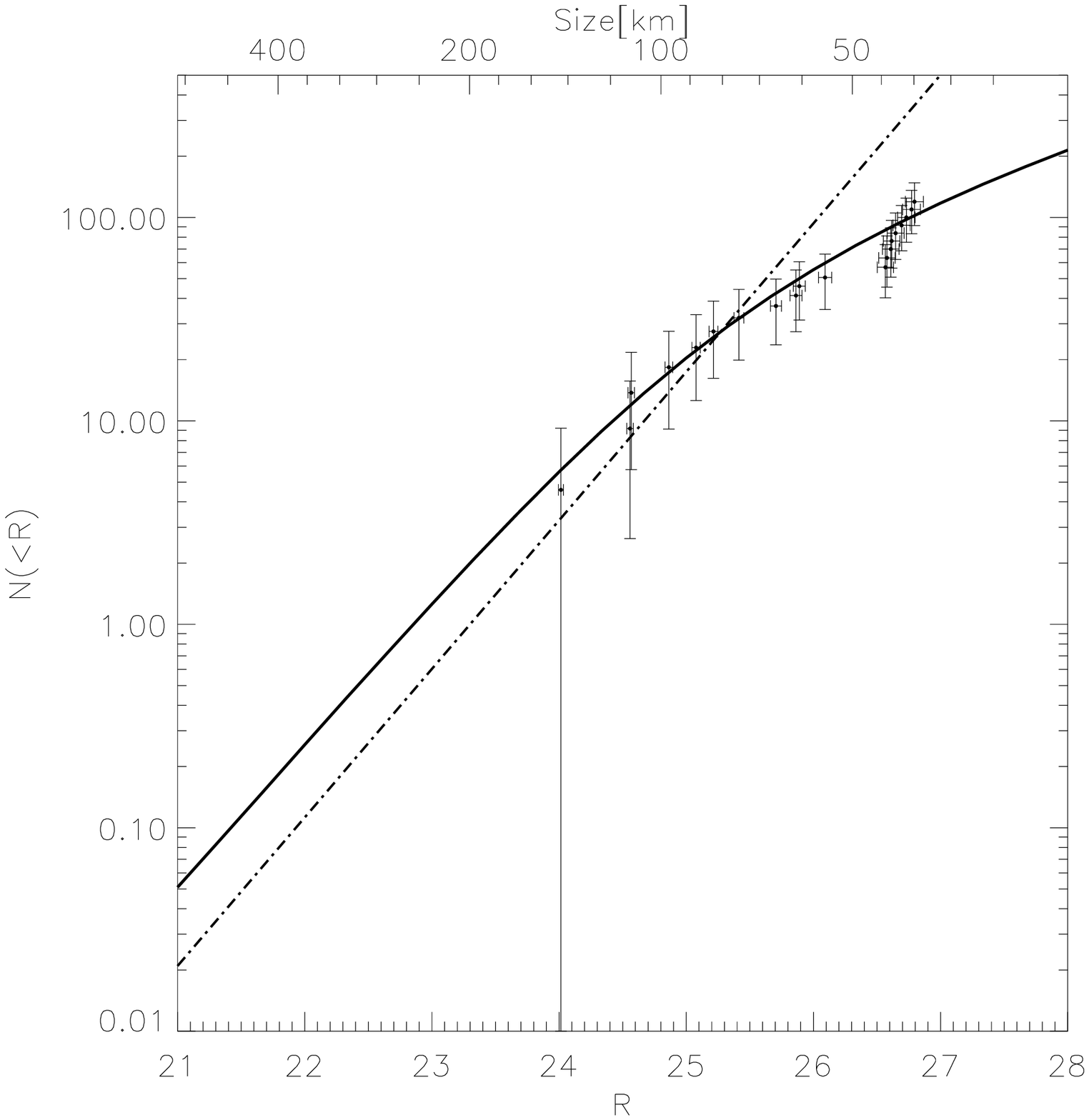}{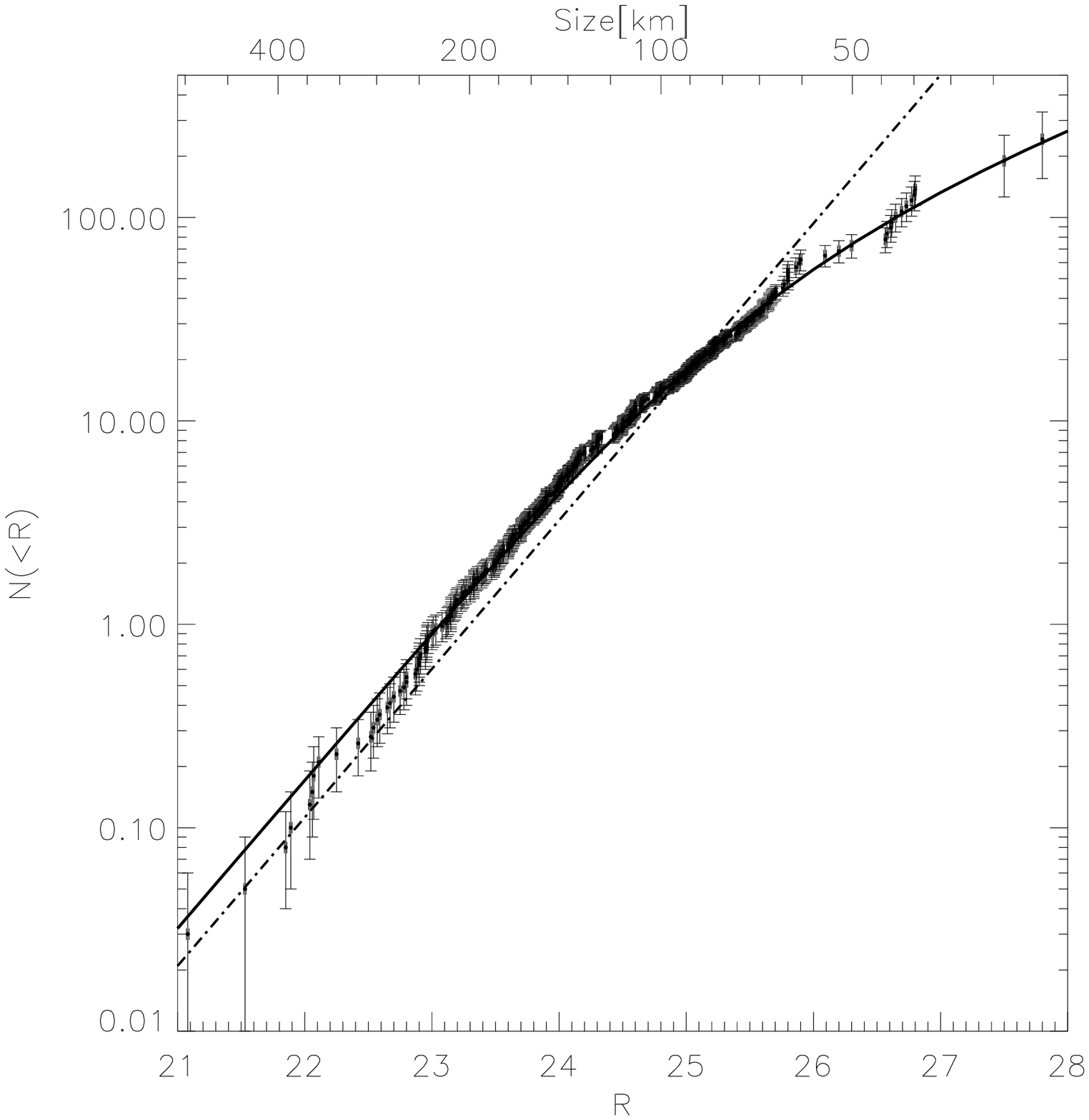}
  \caption{\label{fig:cumfunc} The cumulative number density for the
    objects in our survey is shown on the left panel. The solid line
    is the best previous model \citep{Fuentes.2008}. All surveys
    listed in \citet[Table 2]{Fuentes.2008} plus those included in
    this work are shown in the right panel. The most likely model is
    shown as a solid line (see $\Fig{fig:mcmc}$). The function
    $10^{0.73(R-23.3)}$ is the dot-dashed line. The top axis assumes a
    4\% albedo and every object at $42\au$. The apparent steep slope
    at $R\sim26.7$ is a result of small number statistics and plotting
    the cumulative, rather thatn differential, luminosity function.  }
\end{figure*}

With the standard photometry and efficiency function we construct a
luminosity function for the 20 objects in this field. The result for
this survey is shown on the left panel of Figure
\ref{fig:cumfunc}. The best model for the TNO-number cumulative
function is given by the harmonic mean of two power laws, or double
power law (DPL), as introduced by \citet{Bernstein.2004} and
corroborated by \citet{Fuentes.2008}. The surface number density for
the DPL is given by:
\begin{eqnarray}\label{eqn:sig2}
  \sigma(R) & = & C
   \left[ 10^{-\alpha_1 (R-23)}+10^{(\alpha_2-
   \alpha_1)(R_{eq}-23)-\alpha_2 (R-23)} \right] ^{-1} \nonumber,\\
  C & = & \Sigma_{23} (1+10^{(\alpha_2- \alpha_1)(R_{eq}-23)}) ,
\end{eqnarray}
where $\alpha_1$ and $\alpha_2$ are the exponents for the bright and
faint power law behavior of the model; $\Sigma_{23}$ is the number of
objects expected brighter than $R=23$ and $R_{eq}$ corresponds to the
magnitude at which both power law behaviors meet. Figure
\ref{fig:cumfunc} shows the cumulative surface density
$\Sigma(R)=\int_{-\infty}^R\sigma(x)dx$.

There are not enough objects in this survey alone to constrain the
model; however the data follow the previous best estimate for the
model \citep{Fuentes.2008}. The previous best parameters were
$\alpha_1$=$0.7_{-0.1}^{+0.2}$, $\alpha_2$=$0.3_{-0.2}^{+0.2}$,
$\sigma_{23}$=$2.0_{-0.5}^{+0.5}$ and $R_{eq}$=$24.3_{-0.1}^{+0.8}$,
overplotted in Figure \ref{fig:cumfunc}. Furthermore, this model
predicts 21 detections for this search, consistent with the 20 that
were found.

\begin{figure}[ht]
  \epsscale{1.0} \plotone{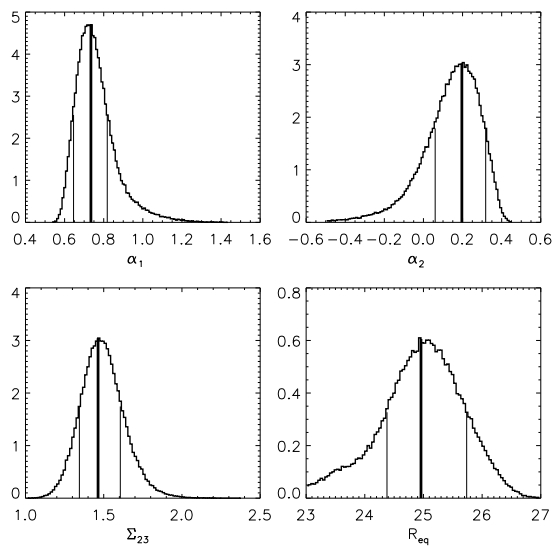}
  \caption{\label{fig:mcmc} The PDF for each model's parameter. The
    most likely value for each parameter and 68\% confidence limits
    are: $\alpha_1$=$0.73_{-0.09}^{+0.08}$,
    $\alpha_2$=$0.20_{-0.14}^{+0.12}$,
    $\sigma_{23}$=$1.46_{-0.12}^{+0.14}$ and
    $R_{eq}$=$25.0_{-0.6}^{+0.8}$. These are plotted as vertical
    lines.  }
\end{figure}

We combined our survey with those listed in \citet[Table
2]{Fuentes.2008}. We only considered objects found at magnitudes at
which the search was over 15\% of the maximum efficiency. All the
objects in our search fulfill this requirement. The total luminosity
function is shown in the right panel of Figure \ref{fig:cumfunc}. The
DPL best fit is determined through a Markov Chain Monte Carlo
simulation (details in \citealt{Fuentes.2008}) with $10^6$ steps, and
an acceptance rate of 25\%. The posterior distribution function (PDF)
for each parameter is plotted in Figure \ref{fig:mcmc}; the most
likely parameters are given by: $\alpha_1$=$0.73_{-0.09}^{+0.08}$,
$\alpha_2$=$0.20_{-0.14}^{+0.12}$,
$\sigma_{23}$=$1.46_{-0.12}^{+0.14}$ and
$R_{eq}$=$25.0_{-0.6}^{+0.8}$.

\section{Conclusions}\label{sec:conc}
We have performed a pencil-beam search of a single Subaru ecliptic
field. The search covered 0.255$\sqdeg$, with a limiting magnitude
$R=26.76$. We found 20 new TNOs with magnitudes between $R=24.0$ and
$26.8$.  As argued by \citet{Bernstein.2004}, it is not surprising
that our faintest detection is near our 50\% efficiency threshold,
given that the luminosity function is much shallower at $R\sim27$.

Including other surveys in the analysis (see \S~\ref{sec:results}) we
derive the most likely DPL model for the luminosity function. With
only 20 new objects there is an important improvement in the model's
PDF from \citet[Figure 13]{Fuentes.2008} to the current result (See
Figure \ref{fig:mcmc}); due to the new detections constraining the
model between $R=26$ and $27$.

The bright end power-law exponent $\alpha_1$=$0.73_{-0.09}^{+0.08}$ is
very close to the results of previous shallower surveys
\citep{Gladman.2001}. The break magnitude
$R_{eq}$=$25.0_{-0.6}^{+0.8}$ is more than 1-$\sigma$ fainter than the
initial estimate by \citet{Bernstein.2004}.  This is also consistent
with earlier surveys reporting excellent fits to a single power model
for magnitudes brighter than $R\sim26$~\citep{Gladman.2001}. Further
data between the current ground-based detection limit of $R\sim27$ and
the HST's at $R\sim28.5$ will determine the break magnitude even more
accurately.

Assuming that all objects are located at an heliocentric distance of
42$\au$, the break in the luminosity function $R_{eq}$ reflects a
break in the size distribution $D$. The corresponding diameter at
which the size distribution breaks is $D=90\pm30(p/0.04)^{-0.5}\km$,
where $p$ is the albedo. Theories predict the existence of a break
\citep{Kenyon.2008}, but at smaller diameters ($D\sim20-40\km$). The
difference reflects the uncertainty in the initial conditions for
numerical simulations and the observational assumptions regarding
physical properties like the albedo.

In this survey both classical and excited TNO populations are
entangled. With better orbital information the size distribution of
each can be studied separately. A comparison with other populations in
the solar system can shed light on their origin.



\end{document}